\documentclass[10.5pt,letterpaper,fleqn, twocolumn]{article}

\usepackage{graphicx}
\usepackage{makeidx}
\usepackage{multicol}
\usepackage{crop}
\usepackage{amstext,amsthm}
\usepackage{amsmath,amssymb}
\usepackage{bm}
\usepackage{natbib}
\usepackage{multirow}
\usepackage{hhline}

\usepackage{multicol}
\usepackage{pslatex}
\usepackage[strict]{changepage}
\newcommand{\be}{\begin{equation}}
\newcommand{\ee}{\end{equation}}
\newcommand{\bea}{\begin{eqnarray}}
\newcommand{\eea}{\end{eqnarray}}
\newcommand{\bne}{\begin{equation*}}
\newcommand{\ene}{\end{equation*}}
\newcommand{\bi}{\begin{itemize}}
\newcommand{\ei}{\end{itemize}}

\newcommand{\bbm}{\begin{bmatrix}}
\newcommand{\ebm}{\end{bmatrix}}

\newcommand{\mr}{\mathrm}

\usepackage[top=0.75in, bottom=0.5in, left=0.83in, right=0.83in]{geometry}
\usepackage[hang,splitrule]{footmisc}





\usepackage{titlesec}		
\usepackage{subcaption}		
\usepackage{indentfirst}	

\newenvironment{myabstract}
{
	\vspace{0.2in}
	\parindent=0cm \textit{Abstract}
	\parindent=0.5cm \hangindent=0.5cm \linebreak
}

\newenvironment{mykeywords}
{
	\vspace{0.2in}
	\parindent=0cm \textit{Keywords}
	
	\parindent=0.5cm
}

\titleformat{\section}
{\normalfont\bfseries}
{}{0.0pt}{\parindent=0cm}[\parindent=0.5cm]

\titlespacing*{\section}
{0pt}{1.5\baselineskip}{0.5\baselineskip}

\titleformat{\subsection}
{\normalfont\itshape}
{}{0.0pt}{\parindent=0cm}[\parindent=0.5cm]

\titlespacing*{\subsection}
{0pt}{1.0\baselineskip}{0.5\baselineskip}

\begin{document}

\twocolumn[{%
	
\phantom\\
\vspace{0.5in}
\begin{center}
\Large{\textbf{Enhanced CNN with Global Features for Fault Diagnosis of Complex \\Chemical Processes}}\\
\end{center}
\vspace{0.2in}

\begin{center}
Qiugang Lu $^{\mr{a,}}$\footnotemark, and Saif S. S. Al-Wahaibi $^{\mr{a}}$ \\

\vspace{0.10in}

$^{\mr{a}}$ Department of Chemical Engineering, Texas Tech University, Lubbock, TX 79409, USA

\end{center}

\begin{myabstract}
Convolutional neural network (CNN) models have been widely used for fault diagnosis of complex systems. However, traditional CNN models rely on small kernel filters to obtain local features from images. Thus, an excessively deep CNN is required to capture global features, which are critical for fault diagnosis of dynamical systems. In this work, we present an improved CNN that embeds global features (GF-CNN). Our method uses a multi-layer perceptron (MLP) for dimension reduction to directly extract global features and integrate them into the CNN. The advantage of this method is that both local and global patterns in images can be captured by a simple model architecture instead of establishing deep CNN models. The proposed method is applied to the fault diagnosis of the Tennessee Eastman process. Simulation results show that the GF-CNN can significantly improve the fault diagnosis performance compared to traditional CNN. The proposed method can also be applied to other areas such as computer vision and image processing. 
\end{myabstract}

\begin{mykeywords}
	Convolutional neural network, Global features, Fault diagnosis, Chemical process.
\end{mykeywords}

\vspace{0.2in}
}]

\footnotetext[1]{\parindent=0cm \small{Corresponding author: Q. Lu (E-mail: {\tt\small jay.lu@ttu.edu}).}}

%
%



\section{1. Introduction}
Deep learning methods have prevailed in various fault diagnosis applications due to their unique effectiveness in automatic feature extractions directly from raw data \cite{zhang2020deep}. Research has been reported on using deep learning methods, e.g., deep belief network \cite{shao2015rolling}, recurrent neural network \cite{jiang2018intelligent}, auto-encoders \cite{zheng2020new}, and convolutional neural networks (CNN) \cite{wen2017new}. Exemplary applications include fault diagnosis of wind turbine gearbox \cite{jing2017convolutional}, rotating machine bearings \cite{zhao2019deep}, and complex chemical processes \cite{huang2022novel}. Such deep learning models can extract abstract features from the data and capture nonlinearity, thereby exhibiting superior performance than traditional machine learning methods \cite{wen2017new}. 

Among the variety of deep learning methods that have been explored, CNN has attracted wide attention due to its excellent performance in complex feature learning and classification \cite{zhang2017new}. CNN was firstly used by \cite{janssens2016convolutional} for fault detection of rotating machinery. 
Further advancements include \cite{wen2017new,zhang2018fault} and the references therein, where a number of strategies were proposed for converting time-series data into images. To capture different levels of features, multi-scale CNN has been put forward where different kernel sizes are proposed, e.g., \cite{chen2021bearing}. 
Along this line, wide kernels have been employed as the first few layers of CNN, followed by small kernels for improving the feature representation \cite{zhang2017new,van2020improved,song2022bearing}. However, these reported studies mainly focus on the fault diagnosis of rotating machinery where the number of variables is small. Study on CNNs and their variants for fault diagnosis of complex chemical processes still remains limited \cite{wu2018deep}. 

Complex chemical processes are featured by high dimensions, with strong spatial and temporal correlations among variables \cite{lu2019fault}. CNN model-based fault diagnosis for chemical processes has only been preliminarily attempted \cite{wu2018deep,huang2022novel}. The obtained diagnosis performance is still far from satisfaction for real-world applications \cite{shao2019multichannel}. Moreover, existing research mainly focuses on extracting \textit{local features} from process data, despite of the efforts on multi-scale CNN to enlarge the receptive field \cite{song2022multi}. For chemical processes, \textit{global features} are also critical due to the complex interconnection of multiple units and thus the widespread coupling between process variables. Specifically, when forming images from multivariate time-series data, variables that are far apart may also possess strong correlations (e.g., see Fig. \ref{fig: Global_features} below). Traditional CNN methods, including multi-scale CNN, cannot directly capture global features and often require deep layers to expand the receptive field to the entire image \cite{song2022multi}. This motivates us to develop a novel CNN architecture that can extract both global and local features in images to improve the fault diagnosis rate. 

In this work, we present a novel global feature-enhanced CNN (GF-CNN) for the fault diagnosis of complex chemical processes. Specifically, in parallel with the convolutional and pooling layers in CNN, we employ a multi-layer perceptron (MLP) network for dimension reduction. That is, the MLP directly maps the vectorized input images  to a low-dimensional feature vector, which is then concatenated with the first fully-connected layer of the CNN. Similar CNN-MLP architecture has been employed in \cite{sinitsin2022intelligent,ahsan2020deep}. However, those works are towards using CNN and MLP to handle multiple input data types (e.g., images and numerical data), instead of using MLP for dimension reduction to assist CNN to capture global features. Moreover, those works focus on bearing fault diagnosis and COVID-19 detection, in contrast to our work on fault diagnosis of complex chemical processes. 

The outline of this paper is as follows. Section \ref{sec: CNN} presents the fundamentals of CNN relevant to this work. Section \ref{sec: methods} describes the proposed GF-CNN in detail, followed by a case study on the Tennessee Eastman process in Section \ref{sec: TEP}. The conclusion is given in Section \ref{sec: conclusion}. 

\section{2. Fundamentals of CNN}
\label{sec: CNN}
For a typical CNN network, convolution and pooling are the two critical operations. In addition, dropout is often employed for regularization to prevent overfitting. 

\subsection{2.1 Convolutional Layers}
For typical convolutional layers, consider a generic 2D feature map $x_{i}^{l}\in\mathbb{R}^{h\times d}$ on layer $l\in\left\{1,\ldots,L\right\}$ and channel $i\in\left\{1,\ldots,N_l\right\}$, where $h$ and $d$ are the height and width of the feature map. The output after a convolutional kernel $k_{i,j}^{l}$ is shown to be \cite{zhao2019deep}
\begin{equation}
	x_{j}^{l+1}=f\left(\sum_{i=1}^{N_l} k_{i,j}^{l} * x_{i}^{l} + b_{j}^{l} \right),~~j=1,\ldots,N_{l+1}, 
\end{equation}
where $x_{j}^{l+1}$ is the $j$-th feature map in layer $l+1$, $*$ is the convolution operation, and $N_{l+1}$ is the number of kernel filters, i.e., the number of channels, in layer $l+1$. Commonly used activation functions $f(\cdot)$ include ReLU, leaky ReLU, or sigmoid function. In this work, we choose to use the ReLU function as the nonlinear activation. When conducting the convolution, a kernel (often square) slides through the entire feature map with a stride $s$. With each stride, the convolution operation above is carried out. In addition, if needed, zero padding can be added to the input map to carry its dimension to the new feature maps. 

\subsection{2.2 Pooling Layers}
Pooling is an operation that is often used after the convolution layers. The purpose of adding the pooling layers is to extract the major features in local regions of the new feature maps after convolutional layers. Therefore, unnecessary details or noise can be filtered out. Moreover, the dimension of the feature maps can be significantly reduced after pooling, and so do the computation time and the number of parameters in the network. Common pooling techniques include average pooling, weighted pooling, and max pooling. In this work, we adopt the max pooling technique \cite{jing2017convolutional}
\begin{equation}
	P_{j}^{l+1} = \max_{x_{j}^{l+1}\in S}~~x_{j}^{l+1},
\end{equation}
where $S$ is pooling block size, $P_{j}^{l+1}$ is the output of the $j$-th feature map in the $(l+1)$-th layer after pooling. The dimension of the feature map is then reduced $S$ times. 

\begin{figure}[tb]
	\centering
	\includegraphics[width=0.40\textwidth]{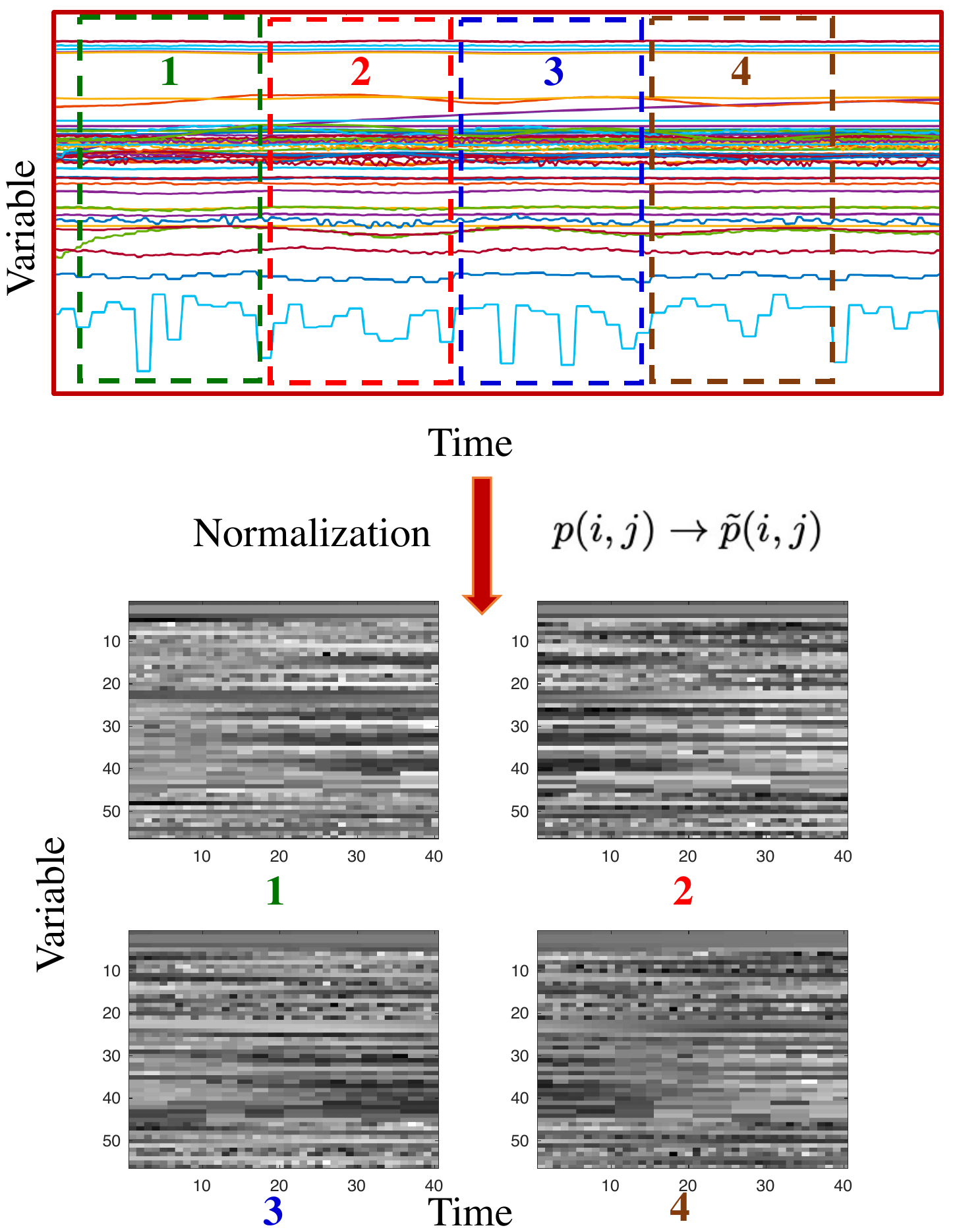}
	\caption{Illustration of signal-to-image conversion. Top: plots of process variables over time; Bottom: the obtained gray images after conversion.}
	\label{fig: signals_to_image}
\end{figure}

\subsection{2.3 Fully-connected (FC) Layers}
The FC layers are located after all convolutional and pooling layers, serving to classify the extracted features from images. Prior to FC layers, the obtained feature maps are flattened into vectors. For the FC part, the output of the $j$-th neuron in the $l$-th layer is
\begin{equation}
x_j^{l} = f\left( \sum_{i=1}^{M_{l-1}} x_{i}^{l-1} w_{i,j}^{l} +  b_{j}^{l} \right),~~j=1,\ldots,M_{l},
\end{equation}
where $M_{l-1}$ and $M_{l}$ are the input and output dimensions for the $l$-th layer, respectively. $w_{i,j}^{l}$ is the weight parameter. The output from the last layer of the FC network is then passed into the softmax function for generating the probability of the input image being classified to each class. 

\subsection{2.4 Dropout}
To prevent overfitting, dropout is a common technique used in CNN. When the dropout is appended to a layer, the output of each neuron will be set to zero with a given probability. For instance, dropout $p_{dp}=0.5$ means that each neuron's output value is set to 0 with probability 0.5. The dropout method can impair the co-adaption of hidden neurons and disable the contribution of those neurons in both backward and forward propagation \cite{wu2018deep}. 

\section{3. Proposed Methods for Fault Diagnosis}
\label{sec: methods}
In this section, we will demonstrate the proposed methods for establishing the GF-CNN model for fault diagnosis.

\subsection{3.1 Signal-to-Image Conversion}
\label{sec: signal-to-image}
For typical chemical processes, the process variables are measured constantly and thus the dataset is in the form of multivariate time series. For 2D-CNN models to consume these data, it is necessary to convert them into images. We will use similar techniques as in \cite{cao2018application,wen2017new} for the signal-to-image conversion. 

Consider a chemical process with $n$ variables and $m$ samples. As shown in Fig. \ref{fig: signals_to_image}, we choose a non-overlapping moving window of width $w$ to slide in the temporal direction. Each sliding window contains $w$ measurements of all the $n$ variables. Then each moving window is considered as an image of dimension $n\times w$ and the measured values of variables are treated as pixels. To ensure the pixel values ranged between $[0,255]$, the standardization below can be performed: 
\begin{align}
\tilde{p}(i,j)=&round\left(\frac{p(i,j)-p_{\min}}{p_{\max}-p_{\min}}\right) \times 255, \nonumber \\
&i=1,\ldots,n,~~j=1,\ldots, w,
\end{align}
where $p(i,j)$ is the $j$-th measurement of the $i$-th variable, and $p_{\min}$ and $p_{\max}$ represent the minimum and maximum signal values of the underlying window, respectively. The pixel intensity at location $(i,j)$ is denoted as $\tilde{p}(i,j)$. Note that the converted image is a gray image with one channel. 

\subsection{3.2 The Proposed GF-CNN Model}
\begin{figure}[tb]
	\centering
	\includegraphics[width=0.45\textwidth]{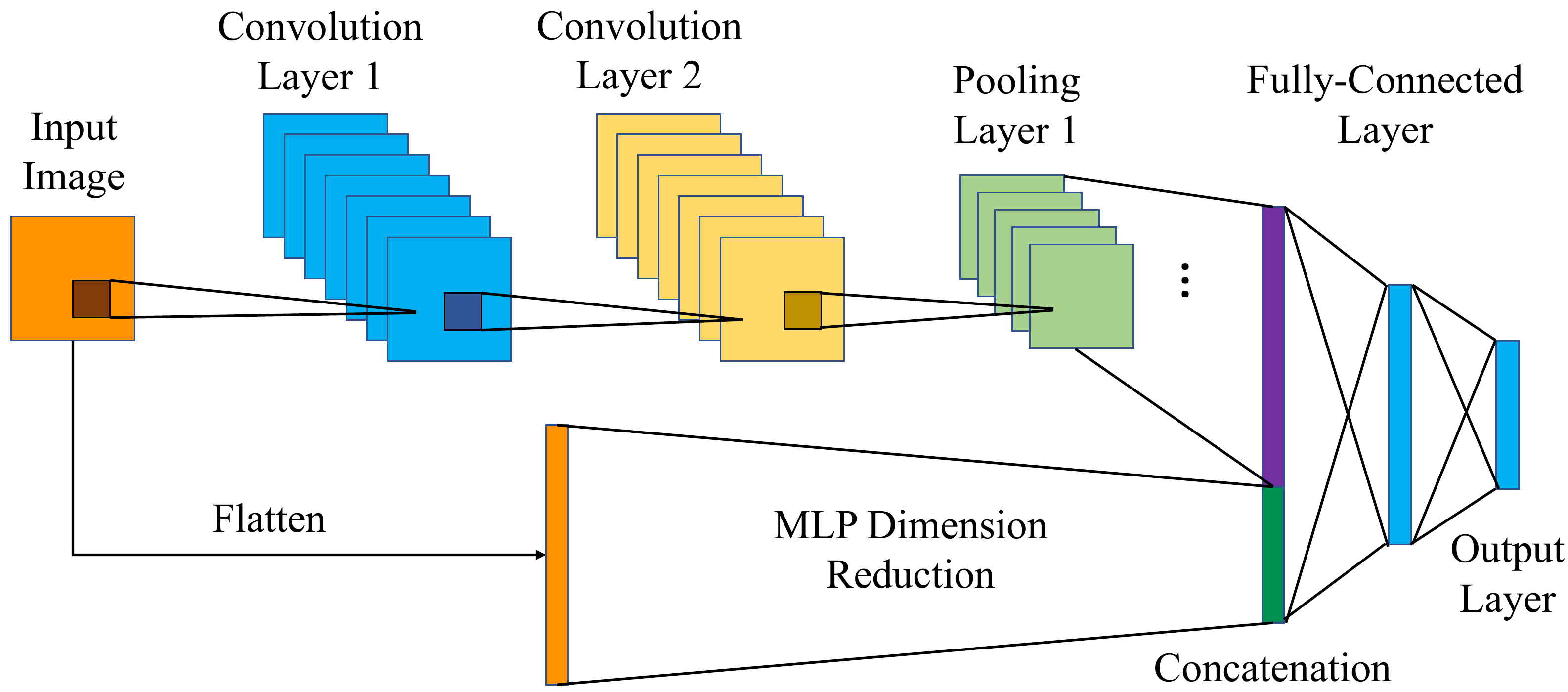}
	\caption{The proposed GF-CNN architecture where the MLP serves to extract global features directly.}
	\label{fig: GF_CNN}
\end{figure}

The proposed GF-CNN combines MLP and CNN for classification so that both global and local features can be extracted from the raw data. Fig. \ref{fig: GF_CNN} shows the structure of the proposed GF-CNN model. Specifically, the top part shows the traditional CNN consisting of multiple convolutional and pooling layers. Denote the mapping from input images to the extracted feature maps in the first FC layer (the purple color in Fig. \ref{fig: GF_CNN}) as:
\begin{equation}
x_{fc,cnn} = f_{\theta_{conv}}\left(x \right),
\end{equation}
where $\theta_{conv}\in\mathbb{R}^{n_{conv}}$ is the collection of training parameters in all convolutional layers. $x \in\mathbb{R}^{n\times w}$ is the input image and $x_{fc,cnn}\in\mathbb{R}^{n_{x,cnn}}$ is the first FC layer feature vector obtained from the CNN. For the bottom part of Fig. \ref{fig: GF_CNN}, we only construct \textit{one layer} of MLP for the purpose of dimension reduction. As a result, the amount of additional parameters to be trained is fairly limited and also far less than that if multiple layers are employed. We further define the mapping from input images to the reduced-order states (the green color in Fig. \ref{fig: GF_CNN}) from the one-layer MLP as:
\begin{equation}
x_{fc,mlp} = f_{\theta_{mlp}}\left(vec(x)\right),
\end{equation}
where $\theta_{mlp}\in\mathbb{R}^{n_{mlp}}$ represents the training parameters from the one-layer MLP network for dimension reduction. $x_{fc,mlp}\in\mathbb{R}^{n_{x,mlp}}$ is the reduced-order state. For the proposed GF-CNN method, vectors $x_{fc,cnn}$ and $x_{fc,mlp}$ are concatenated as the first layer of the FC network. The FC network afterwards can be represented mathematically as
\begin{equation}
\hat{y} = f_{\theta_{fc}}\left(x_{fc}^{1}\right),~~x_{fc}^{1}=[x_{fc,cnn}^{\top}~~ x_{fc,mlp}^{\top}]^{\top}\in\mathbb{R}^{n_{x,cnn}+n_{x,mlp}},
\end{equation}
where $\hat{y}\in\mathbb{R}^{C}$ ($C$ is the number of classes) is the output of the GF-CNN and $\theta_{fc}\in\mathbb{R}^{n_{fc}}$ is the training parameter associated with the FC layers. $x_{fc}^{1}$ is the concatenation of the flattened feature map from CNN and reduced-order state from MLP. 

With the GF-CNN model above, the overall training problem can be formulated as (cross-entropy criterion):
\begin{equation}
\min_{\theta}~~ J(\theta)=-\frac{1}{N}\left[\sum_{i=1}^{N}\sum_{c=1}^{C} \mathbb{I}(y_{i}=c)\log \mathbb{P}(\hat{y}_{i}=c|x_i;\theta)\right],
\end{equation}
where $\theta=\{\theta_{conv},\theta_{mlp},\theta_{fc}\}$ gathers all training parameters from each block; $y_i$ is the one-hot true label for the input image $x_i$; $\mathbb{P}(\hat{y}_{i}=c|x_i;\theta)$ is the predicted probability of classifying $x_i$ into class $c$ from the softmax function; $N$ is the total number of training samples; and $\mathbb{I}(\cdot)$ is the indicator function. 

\subsection{3.3 Model Complexity}
\label{sec: model_complexity}
Incorporating a single-layer MLP as the dimension reduction into CNN introduces additional parameters. Here we will show that such a GF-CNN network will not significantly increase the number of parameters (i.e., model complexity) compared with the pure CNN. In general, for process data, the number of dominant features $n_{x,mlp}$ is usually much less than the data dimensionality \cite{lu2019fault}. That is, $n_{x,mlp} \ll n\cdot w$ and usually we also have $n_{x,mlp}\ll n_{x,cnn}$. Note that the parameters in all FC layers $n_{fc}$ consist of those induced by CNN $n_{fc,1}$ and those by single-layer MLP $n_{fc,2}$, i.e.,  $n_{fc}=n_{fc,1}+n_{fc,2}$. From the argument above that $n_{x,mlp}\ll n_{x,cnn}$, we thus have $n_{fc,2}\ll n_{fc,1}$. In addition, since we only utilize one-layer MLP for dimension reduction, it can be inferred that, in general, the total number of induced parameters from the MLP is much less than that from the CNN: $n_{mlp} + n_{fc,2} \ll n_{conv} + n_{fc,1}$. In other words, adding the global feature extraction will not introduce many parameters. This observation will be verified in the case study below. In summary, compared with traditional CNN, the GF-CNN can boost the performance (will be shown below) by only slightly increasing the model complexity.

\begin{figure}[tb]
	\centering
	\includegraphics[width=0.45\textwidth]{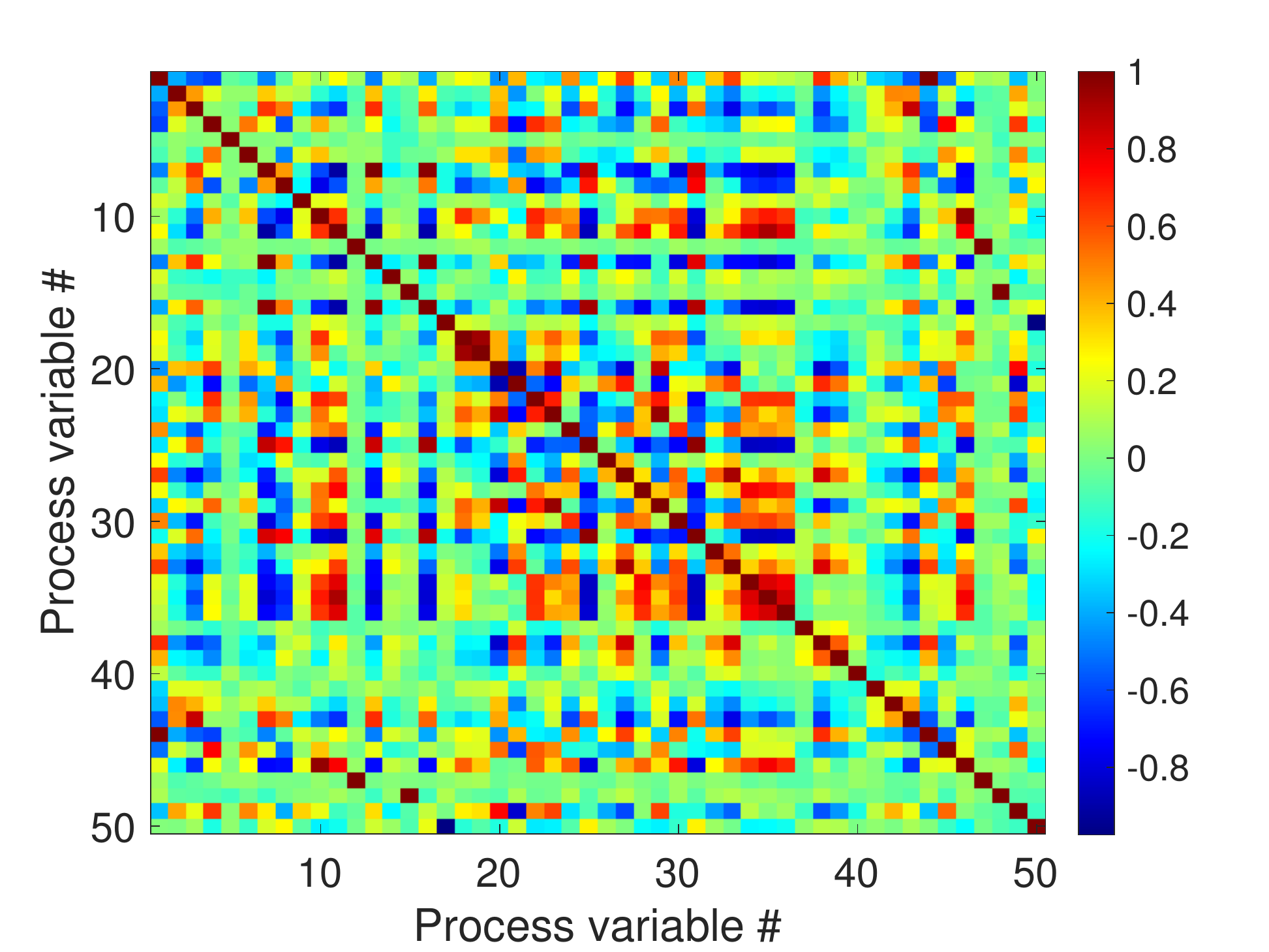}
	\caption{The correlation coefficient colormap for all 50 process variables from TEP.}
	\label{fig: Global_features}
\end{figure}

\section{4. Case Study on the Tennessee Eastman Process}
\label{sec: TEP}
We will use the benchmark Tennessee Eastman process (TEP) as the platform for validating the fault diagnosis performance of our GF-CNN model. The TEP simulates a complex chemical process consisting of five units: a two-phase reactor, a vapor/liquid extractor, a condenser, a compressor, and a stripper. The closed-loop operation data of TEP contains 52 variables, including 41 process measurements and 11 manipulated variables. Normal and different types of faults are programmed into the simulator to generate the corresponding dataset. In this work, we use the large-sample TEP data available from Harvard University \cite{DVN/6C3JR1_2017}. 

\subsection{4.1 Data Description and Global Features}
The selected dataset contains 20 different types of faults. Similar to \cite{wu2018deep}, we drop XMV(5) (compressor recycle valve) and XMV(9) (stripper steam valve) from the data due to their constant values in some simulations. As a result, the number of process variables $n=50$. The sampling period is 3 min. For the training data, each fault class is simulated for 24 hours under 40 random seeds, which gives a total number of  19,200 samples. For the dataset of each fault, we choose the moving window size $w=20$ without overlapping, yielding 960 gray-scale images. The total number of images for all 20 faults is 19,200. For the test data, each fault is simulated for 40 hours under 7 different random seeds. The window size is also $w=20$ and this gives us 5,600 samples of data, or equivalently 280 images for each fault. Thus, the total number of test images for all 20 faults is 5,600. 

The colormap in Fig. \ref{fig: Global_features} shows the correlation coefficients among all 50 variables, corresponding to the 50 rows of each image. It is observed that not all strong correlations exist around the diagonal line. Instead, many strong correlations are dispersed across the entire colormap. In other words, different from traditional images where the textures are often localized, for the images converted from multivariate time series, \textit{non-local features} are common and sometimes dominant. In fact, the order of process variables into the rows of the image can make two variables strongly correlated, although they are far apart. 

\subsection{4.2 Data Preprocessing}
Prior to the signal-to-image conversion, both training and test data are normalized. Specifically, for the training data, the overall mean  $\bar{x}_{tr}\in\mathbb{R}^{n}$ and standard deviation  $\sigma_{tr}\in\mathbb{R}^{n}$ vectors are computed. Each training and test data sample is normalized as below:
\begin{equation}
\tilde{x}_{tr}^{i} = \frac{x_{tr}^{i}-\bar{x}_{tr}}{\sigma_{tr}}, \quad \tilde{x}_{te}^{i} = \frac{x_{te}^{i}-\bar{x}_{tr}}{\sigma_{tr}}.
\end{equation}
Note that here we are using the mean and standard deviation from the training data for normalizing the test data. The purpose is to keep the separability between different faults; Otherwise, some separating features may be lost if the test data uses its own mean and standard deviation for normalization. After normalization, both training and test data will be passed to the signal-to-image conversion in Section \ref{sec: signal-to-image} to obtain images.  

\subsection{4.3 Model Design and Training}
To comprehensively assess the fault diagnosis performance of the proposed method, we design 6 model architectures with increasing complexity for CNN and GF-CNN, shown in Table \ref{table: Models}. For GF-CNN models, the reduced-order feature vector from the MLP has a dimension $n_{x,mlp}=10$, much less than the dimension $n_{x,cnn}$ from the last feature map of the CNN. Consequently, each GF-CNN model only has slightly more parameters than the corresponding CNN model, consistent with the analysis in Section \ref{sec: model_complexity}. Each model is trained on the training images and validated against the test images. The selected hyperparameters are shown in Table \ref{table: Hyperparameters} below. All models are trained on the NVIDIA Tesla V100 GPU with 16 GB memory. 
\begin{table}[h]
	\centering
	\caption{Selected hyperparameters for training the models}
	\label{table: Hyperparameters}	
	\begin{tabular}{ll|ll}
		\hline
	Hyperparameter  &  Value  & Hyperparameter  &  Value \\ \hline
	Batch size: & 128 &	Epochs: & 50 \\
	Learning rate: & 0.001 & Dropout rate: & 0.5 \\
	\hline	
	\end{tabular}%
\end{table}

\begin{table*}[h]
	\centering
	\caption{The designed CNN and GF-CNN architectures for the case study}
	\label{table: Models}	
	\begin{tabular}{c|l|r|l|r}
		\hline
		\multirow{2}{*}{Models}	& \multicolumn{2}{c|}{CNN} &  \multicolumn{2}{c}{GF-CNN} \\
		\hhline{~----}
				 & Structure & Parameter \#  & Structure & Parameter \# \\  \hline
		\# 1 & C(16)-P(2)-F(100)*  & 347,880 & C(16)-P(2)-G(10)-F(100)* & 358,890 \\ \hline
		\# 2 & C(16)-P(2)-C(32)- P(2,1)-F(300)*   &  644,720   & C(16)-P(2)-C(32)-P(2,1)-G(10)-F(300)* &  657,730  \\ \hline
		\# 3 & C(32)-P(2)-C(64)-P(2,1)-F(300)*  &  1,292,336 & C(32)-P(2)-C(64)-P(2,1)-G(10)-F(300)* & 1,305,346  \\  \hline
		\multirow{2}{*}{\# 4} & C(64)-P(2)-C(64)- &  \multirow{2}{*}{1,653,744} & C(64)-P(2)-C(64)-C(128)- & \multirow{2}{*}{1,666,754}  \\
		& C(128)-P(2,1)-F(300)*  & &  P(2,1)-G(10)-F(300)* & \\ \hline
		\multirow{2}{*}{\# 5} & C(32)-C(64)-P(2)- &  \multirow{2}{*}{2,518,192} &  C(32)-C(64)-P(2)-C(128)- & \multirow{2}{*}{2,531,202}  \\
		& C(128)-P(2,1)-F(300)* & &  P(2,1)-G(10)-F(300)* & \\ \hline
		\multirow{2}{*}{\# 6} & C(64)-C(64)-P(2)-C(128)- &  \multirow{2}{*}{3,331,312} &  C(64)-C(64)-P(2)-C(128)-C(256)-  & \multirow{2}{*}{3,344,322}  \\
		& C(256)-P(2,1)-F(300)* & &  P(2,1)-G(10)-F(300)* & \\ 
		\hline		
		\multicolumn{5}{l}{C($n$): Convolution with $n$ kernels of $3\times 3$; P($n$): $n\times n$ max pooling;  P($n$,$m$): $n\times m$ max pooling; } \\
		\multicolumn{5}{l}{F($n$): FC layer with $n$ neurons; G($n$): Global feature of dimension $n$ from MLP; *: Dropout with rate $p_{dp}=0.5$.}
	\end{tabular}%
\end{table*}

\subsection{4.4 Results}
In this subsection, we will demonstrate the fault diagnosis performance of the proposed method for the TEP, and also compare it with the standard CNN without directly considering global features. For each model, we repeat the training and testing 10 times to account for the randomness from the optimization algorithm and data shuffling. The fault diagnosis rate (FDR), for the $i$-th class, is defined as
\begin{equation}
FDR = \frac{P}{P+B},
\end{equation}
where $P$ represents the number of image samples correctly classified into the $i$-th class, $B$ is the number of image samples classified into the other classes, and $P+B$ is the total number of images in class $i$, i.e., the ground truth. 

\begin{figure}[tb]
	\centering
	\includegraphics[width=0.5\textwidth]{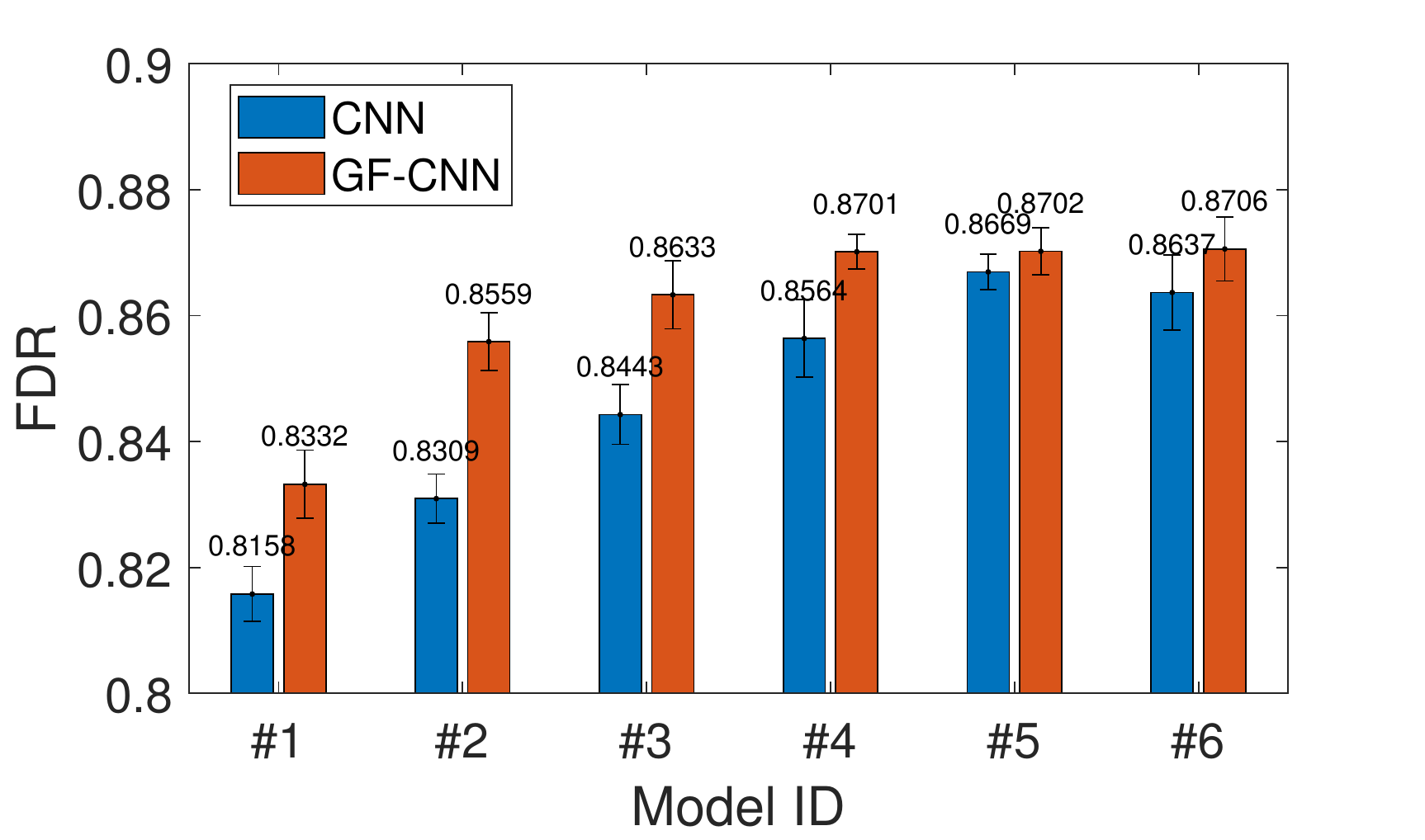}
	\caption{Fault diagnosis performance of the six designed CNN (blue) and GF-CNN (red) models on the TEP data.}
	\label{fig: Model_accuracy}
\end{figure}

 Fig. \ref{fig: Model_accuracy} shows the comparison of the overall FDR between 6 different CNN and GF-CNN models averaged over 20 faults. It is observed that for simple model architectures (e.g., model \#1, \#2, and \#3), the GF-CNN outperforms the CNN significantly, with a slight increase in the number of learnable parameters. In addition, a simple GF-CNN can achieve a comparable or even better performance than a complex CNN model. For instance, GF-CNN model \#2 can yield comparable performance to CNN models \#4, but the number of parameters is far less (657,730 vs. 1,653,744). This is because CNN usually requires deep layers to make the receptive field to cover the entire image. Thus, shallow CNN may not capture global features. However, for GF-CNN, it can directly acquire global features without unnecessarily building deep layers. For deeper models (e.g., model \#5, \#6), the performance difference between CNN and GF-CNN becomes less apparent because deep CNN can extract global features. 

\begin{figure}[bt]
	\centering
	\includegraphics[width=0.48\textwidth]{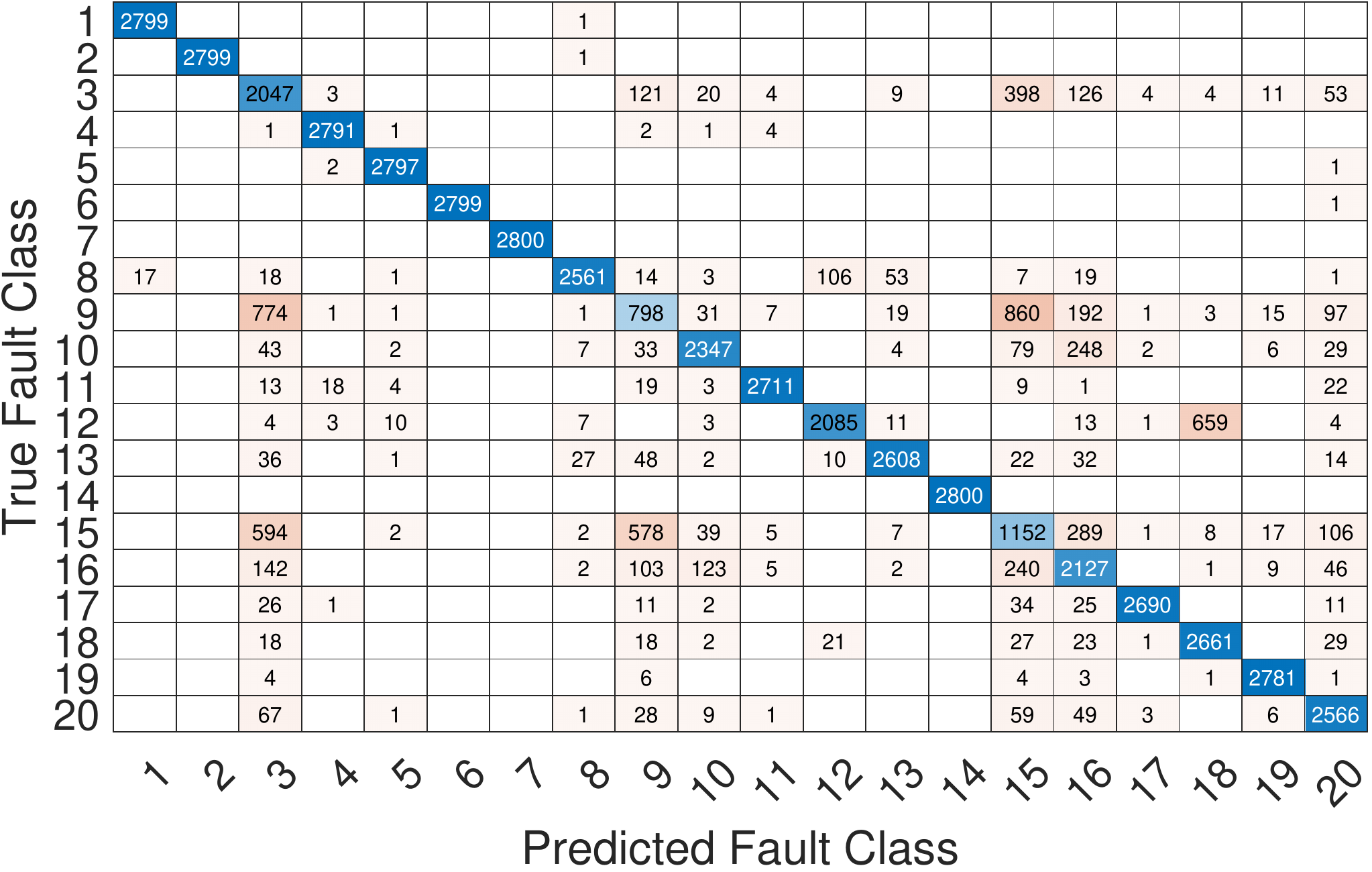}
	\caption{The confusion matrix of the fault diagnosis based on the FG-CNN model \#4.}
	\label{fig: Confusion_MLP_CNN}
\end{figure}

Among those model architectures, we select GF-CNN model \#4 and CNN model \#5 for further comparison. Note that GF-CNN model \#4 has much less trainable parameters than CNN model \#5. Table \ref{table: Results} shows the FDR across all 20 faults averaged from 10 runs. It is clear that for most faults, the GF-CNN yields better fault detection rates, except for Faults 9, 12, and 16, where CNN gives a slightly higher FRD. In particular, for Faults 3, 8, 13, 15, and 20, the GF-CNN outperforms the CNN significantly. Fig. \ref{fig: Confusion_MLP_CNN} shows the confusion map of fault diagnosis based on GF-CNN model \#4, where most faults can be correctly diagnosed. It also indicates that Faults 3, 9, 15, and 18 are difficult to classify, consistent with the observations in literature \cite{lu2019fault}. These results validate the effectiveness of GF-CNN in improving FDR with a much simpler structure than CNN. 

\begin{table}[tb]
	\centering
	\caption{FDR of GF-CNN model \#4 and CNN model \#5}
	\label{table: Results}	
	\begin{tabular}{lll|lll}
		\hline
		Fault &  CNN & GF-CNN  & Fault & CNN  & GF-CNN  \\ \hline
		1 & 0.9989 & \textbf{0.9996} & 11& 0.9671 & \textbf{0.9682} \\
		2 & 0.9921 & \textbf{0.9996} & 12 & \textbf{0.7911} & 0.7446 \\
		3 & 0.6700 & \textbf{0.7311} & 13 & 0.9279 & \textbf{0.9314} \\
		4 & \textbf{0.9971} & 0.9968 & 14 & 1.0000 & 1.0000 \\
		5 & 0.9968 & \textbf{0.9989} & 15 & 0.3993 & \textbf{0.4114} \\
		6 & \textbf{1.0000} & 0.9996 & 16 & \textbf{0.7657} & 0.7596 \\
		7 & 0.9993 & \textbf{1.0000} & 17 & 0.9607 & 0.9607 \\
		8 & 0.9096 & \textbf{0.9146} & 18 & 0.9464 & \textbf{0.9504} \\
		9 & \textbf{0.3014} & 0.2850 & 19 & 0.9893 & \textbf{0.9932} \\
		10 & 0.8300 & \textbf{0.8382} & 20 & 0.8961 & \textbf{0.9197} \\
		\hline	
	\end{tabular}%
\end{table}

\section{5. Conclusion}
\label{sec: conclusion}
This article presents a novel global feature-enhanced CNN with an application to the fault diagnosis of complex chemical processes. For the proposed method, a simple single-layer MLP network is utilized to reduce the dimension of input images and extract global features. Such features are then integrated with the local features obtained from CNN to improve the classification performance. Different architectures of CNN and GF-CNN are designed and validated through the benchmark Tennessee Eastman process dataset. Fault diagnosis results reveal that the GF-CNN with a simple structure can greatly improve the fault diagnosis performance by accounting for global features. The proposed method can be easily extended to other areas such as computer vision and image processing. Future work includes combining multi-scale CNN with MLP to obtain both global and different levels of local feature representation.

\section{Acknowledgment}
Q. Lu acknowledges the startup funds from Texas Tech University. S. Al-Wahaibi acknowledges the Distinguished Graduate Student Assistantships from Texas Tech University.


%
%

\bibliographystyle{chicago}
\section{References}


\def\refname{}
\def\bibsection{}

\bibliography{references.bib}%


%
%
%
%
%


\end{document}